\documentclass[aps,prb,showpacs]{revtex4}
\usepackage{graphicx}
\usepackage{amsmath}

\begin{document}
\title{Elastic and Chemical Contributions to the Stability of Magnetic Surface Alloys on Ru(0001)}
\author{Madhura Marathe, Mighfar Imam and Shobhana Narasimhan}
\affiliation{Theoretical Sciences Unit, Jawaharlal Nehru Centre for Advanced Scientific Research, 
Jakkur, Bangalore-560064, India.}

\begin{abstract}
We have used density functional theory to study the structural stability of surface alloys. Our systems
consist of a single pseudomorphic layer of $M_xN_{1-x}$ on the Ru(0001) surface, 
where $M$ = Fe or Co, and $N$ = Pt, Au, Ag, Cd, or Pb. Several of the combinations studied by us display
a preference for atomically mixed configurations over phase-segregated forms.
We have also performed further {\it ab initio} calculations to obtain the parameters describing the elastic
interactions between atoms in the alloy layer, including the effective atomic sizes at the surface. 
We find that while elastic interactions favor alloying for all the systems considered by us, in some
cases chemical interactions disfavor atomic mixing. We show that a simple criterion (analogous to the Hume-Rothery
first law for bulk alloys) need not necessarily work for strain-stabilized surface alloys, because of the
presence of additional elastic contributions to the alloy heat of formation, that will tend to oppose phase segregation. 
\end{abstract}


\maketitle
\section{Introduction}
It has been known since ancient times that alloying two metals can give rise to a new material with
properties that are improved over those of the constituent metals. For example, alloys can have superior
mechanical or magnetic properties, an increased resistance to corrosion, or constitute good catalysts. However, not
all pairs of metals form stable alloy phases. The rules governing alloy formation in the bulk were first
formulated by Hume-Rothery.\cite{Hume-Rothery} The first of these empirical laws states that if the atomic size
mismatch is greater than 15\%, phase segregation is favored over the formation of solid solutions. Thus, many
pairs of metals are immiscible (or nearly so) in the bulk.

In recent years, it has become apparent that surface science can extend the chemical phase space available for
the search for new alloy systems. It has long been known that bulk alloys exhibit surface segregation, so that
the chemical composition at the surface can differ considerably from that in the bulk. However, the field of
surface alloying gained additional interest when it was discovered that even metals that are immiscible in
the bulk can form stable surface alloys as a result of the altered atomic environment at
the surface.\cite{Nielsen, Roder, Neugebauer} 
These alloys display atomic mixing that is confined to the surface layer or, in
some cases, the top few layers. These results were explained by Tersoff,\cite{Tersoff} who argued that
in cases where there is a large size mismatch, as a result of which the energetics are dominated by strain effects,
alloying will be disfavored in the bulk but favored at the surface.

Subsequently, another class of surface alloys has emerged, where two metals that differ in size are co-deposited on a third metal of intermediate size. In such systems, any single-component pseudomorphic layer will be under tensile or compressive stress (that may or may not be relieved by the formation of dislocations);\cite{AgonRu}
however if the two elements were to mix, the stress would presumably be considerably relieved.
Thus, the strain imposed by the presence of the substrate promotes alloying
in the surface layer. Some examples of such strain-stabilized surface alloys are an Ag-Cu monolayer 
on Ru(0001),\cite{AgCuonRu,Schick,Schick1} Pd-Au/Ru(0001),~\cite{PdAuonRu} and Pb-Sn/Rh(111).\cite{PbSnonRh} 

Hitherto, the guiding principle in the search for such systems has been the rule-of-thumb that the (bulk)
nearest-neighbor (NN) distance of the substrate should be the average of the NN distances of the two overlayer
elements. However, this simple criterion does not necessarily work. For example, Thayer {\it et al.} have
studied the Co-Ag/Ru(0001) system.\cite{Thayer, Thayer1} At first sight, this system would seem to be a good candidate for the formation of a strain-stabilized surface alloy, since the NN distance for Ag is larger than that
of Ru by 8\%, while that of Co is smaller by 7\%. However, instead of forming an atomically mixed structure,
it was found that the stable structure consisted of Ag droplets surrounded by Co. After doing a combined
experimental and theoretical study, these authors concluded that chemical bonding between Ag and Co is
disfavored in this system, and the observed structure results from a lowering of stress at the boundary between
Co and Ag islands.

In this paper, we examine ten different bi-metallic systems on a Ru(0001) substrate. Some of the questions
that we hope to address include: (i) is it only the mean size of the overlayer atoms that matters, or do individual sizes also matter? (ii) can one develop a criterion based on atomic size that will predict
whether or not a surface alloy will form? (iii) how different are atomic sizes at the surface compared to those
in the bulk?, and (iv) what is the relative importance of elastic and chemical interactions? 

The bi-metallic systems we have considered all consist of one magnetic metal $M$ (Fe or Co) and one non-magnetic
metal $N$ (Pt, Ag, Au, Cd or Pb), co-deposited on Ru(0001) to form a surface alloy of the form  $M_xN_{1-x}/S$,
where $S$ denotes the Ru substrate. Such systems, involving one magnetic and one non-magnetic element, are of interest because alloying can, in some cases, improve magnetic properties. Conversely, in some applications,\cite{Tober, Yang} one would prefer that instead of mixing at the atomic level, the system should spontaneously organize into a pattern consisting of alternating domains of the magnetic and the non-magnetic element. Ru(0001) was chosen as
the substrate, in part because of its intermediate NN distance, and in part because its hardness and immiscibility
with the other elements make it less likely that the alloy elements will penetrate into the bulk.  The bulk NN distances,
$a_M$, of the two magnetic metals Fe and Co, are
about 7-8\% less than $a_S$, the NN separation in the Ru substrate, while all five non-magnetic metals we have considered have bulk NN distances, $a_N$, larger than that of Ru. However, the $N$ metals we have chosen display a large variation in size: the NN distance in Pt is approximately 3\% more than that in Ru, while in Pb the discrepancy is  26\%. Accordingly, only Fe-Pt and Co-Pt fall within the 15\% range of the Hume-Rothery criterion for bulk alloys;
alloys of Fe and Co with Au and Ag fall slightly outside this range, while those with Cd and Pb fall well
outside the range. If there is a size-dependent trend that determines whether or not alloying is favored,
then one might hope that it will show up upon examining these ten systems. In Table I, we have given the
average NN separation, $(a_M + a_N)/2$, using experimental values for the bulk metals. Upon examining how
close these values lie to $a_S$ = 2.70 \AA, one might expect (using the simple criterion mentioned above)
that Fe-Au, Fe-Ag, Co-Au and Co-Ag might be good candidates for forming strain-stabilized surface alloys,
and Fe-Cd, Co-Cd, Fe-Pt and Co-Pt may be possibilities, but Fe-Pb and Co-Pb surface alloys should be highly
unlikely to form. As we will show below, these simple-minded expectations are not necessarily borne out.

Of the ten systems we consider in this paper, we are aware of previous studies on only two of them: Co-Ag/Ru(0001),\cite{Thayer, Thayer1} and Fe-Ag/Ru(0001).\cite{Yang} In both these cases, it was found that chemical interactions
dominate over elastic ones, and the atomically mixed phase is disfavored.

\begin{table}[tb]
\begin{center}
\begin{tabular}{|c|c|c|c|c|c|}
\hline
\ & Pt & Au & Ag & Cd & Pb \\
\hline
Fe & 2.63 & 2.69 & 2.69 & 2.73 & 2.99 \\
\hline
Co & 2.64 & 2.70 & 2.70 & 2.75 & 3.01 \\
\hline
\end{tabular}
\end{center}
\caption{Values in \AA \ of $(a_M+a_N)/2$, the average of the nearest neighbor (NN) spacings of $M$ and $N$ in their bulk structures.
By comparing these numbers with $a_S$, the NN distance of Ru in the bulk = 2.70 \AA, one expects Co-Au and Co-Ag to form stable alloys, 
and Pb alloys to be unstable. In this table, all the values used are experimental values, taken from Ref.~\onlinecite{AscMer}.}\label{tab-bulk-avg}
\end{table}

\section{Computational details}
All the calculations are done using \textit{ab initio} spin polarized density functional theory with the PWscf package 
of the Quantum-ESPRESSO distribution.~\cite{pwscf} A plane-wave  basis set is used with a kinetic energy cutoff of 20 Ry. 
The charge-density cutoff value is taken to be 160 Ry. Ultrasoft pseudopotentials~\cite{uspp} are used to describe the interaction between ions and valence electrons. 
For the exchange correlation functional, a Generalized Gradient Approximation (GGA) of 
the Perdew-Burke-Ernzerhof form~\cite{pbe} is used.
As all the systems are metallic, the Methfessel-Paxton smearing technique~\cite{mp} is used with 
the smearing width equal to 0.05 Ry.

Convergence with respect to the basis size and the k-point grid has been carefully verified. For the bulk structure calculations, we have used the common crystal phase of each element. 
The k-points used for Brillouin zone integrations form an 8$\times$8$\times$8 Monkhorst-Pack grid\cite{grid}
for bulk calculations, and a 4$\times$4$\times$1 grid for surface calculations.
To study the surface properties, the supercell approach is used, with a unit cell that
includes a slab and some vacuum layers. The slab used corresponds to a $2 \times 2$ surface unit cell, and contains six Ru layers 
to model the substrate. Our results for the energetics were obtained
with one alloy overlayer (deposited on one side of the substrate) 
and seven vacuum layers (approximately 17.4 \AA); we have allowed the alloy overlayer and the three topmost layers of Ru to relax, 
using Hellmann-Feynman forces. However, when performing calculations to see how the surface stress
of monolayers of $M$ or $N$ on $S$ varied with in-plane distance, the monolayer was deposited symmetrically
on both sides of the slab, and the central layers of the slab were held fixed, while the outer layers on
both sides were allowed to relax.
  
\begin{figure}[tb]
\centering
\includegraphics[scale=0.15]{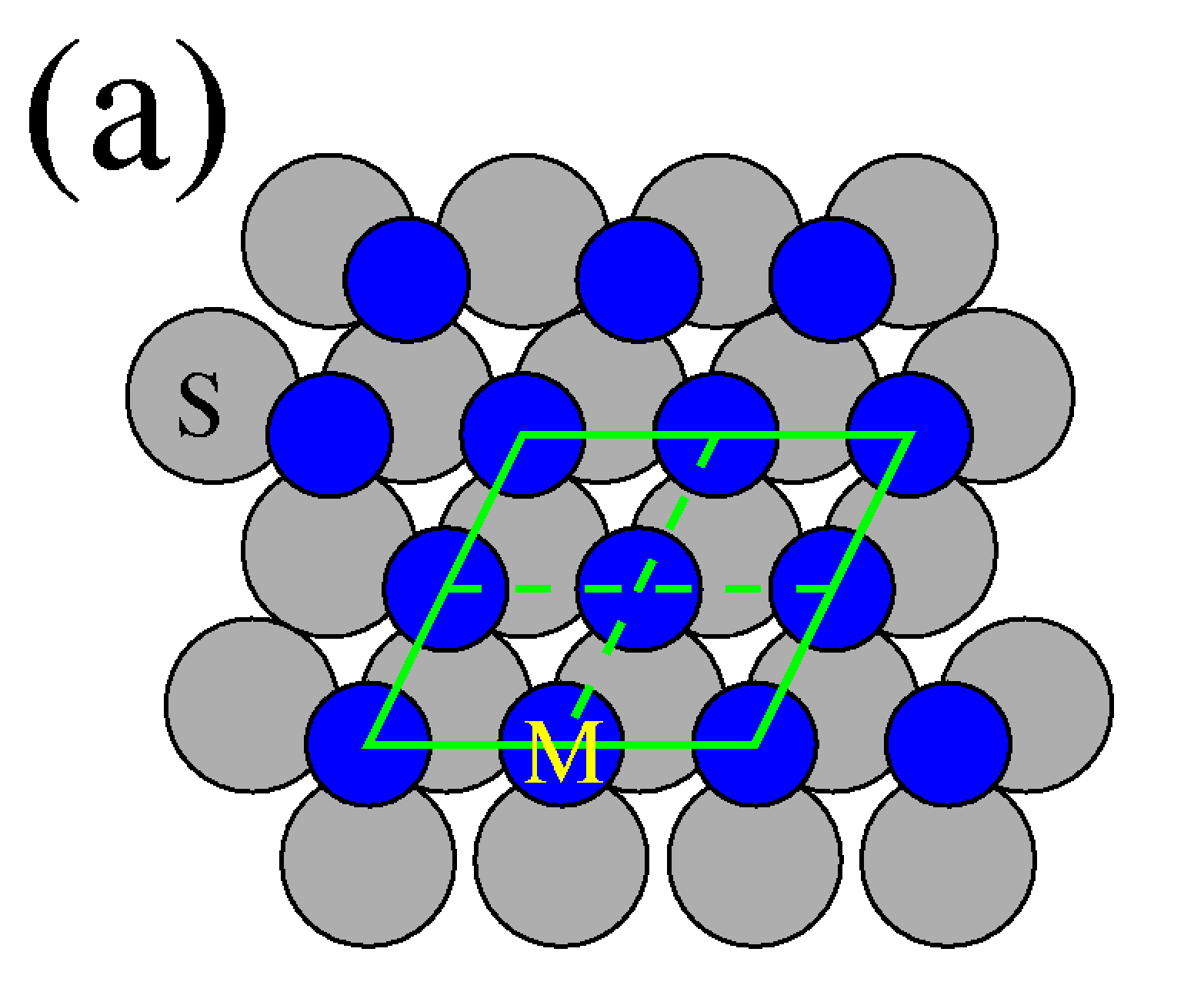}
\includegraphics[scale=0.15]{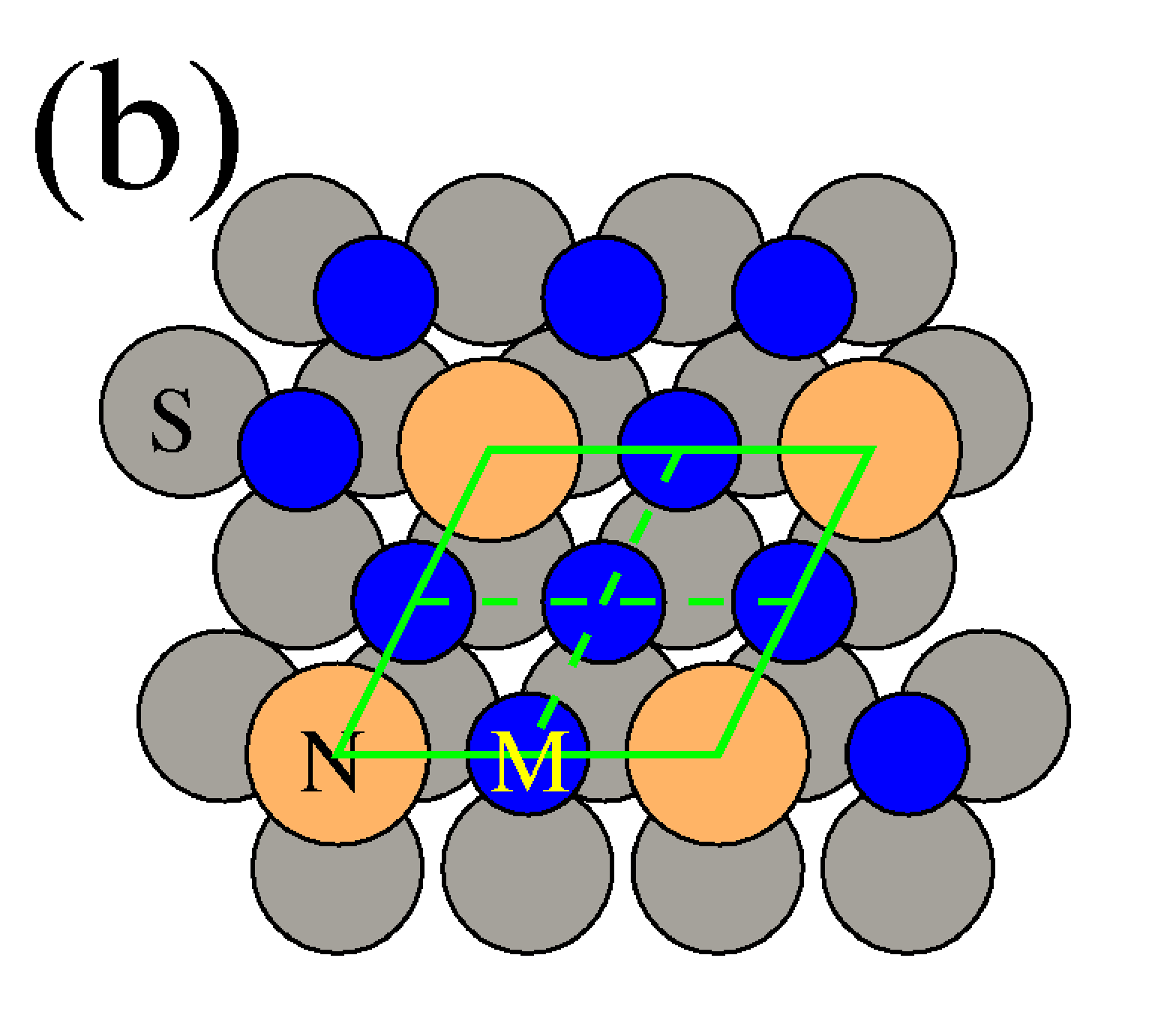}
\includegraphics[scale=0.15]{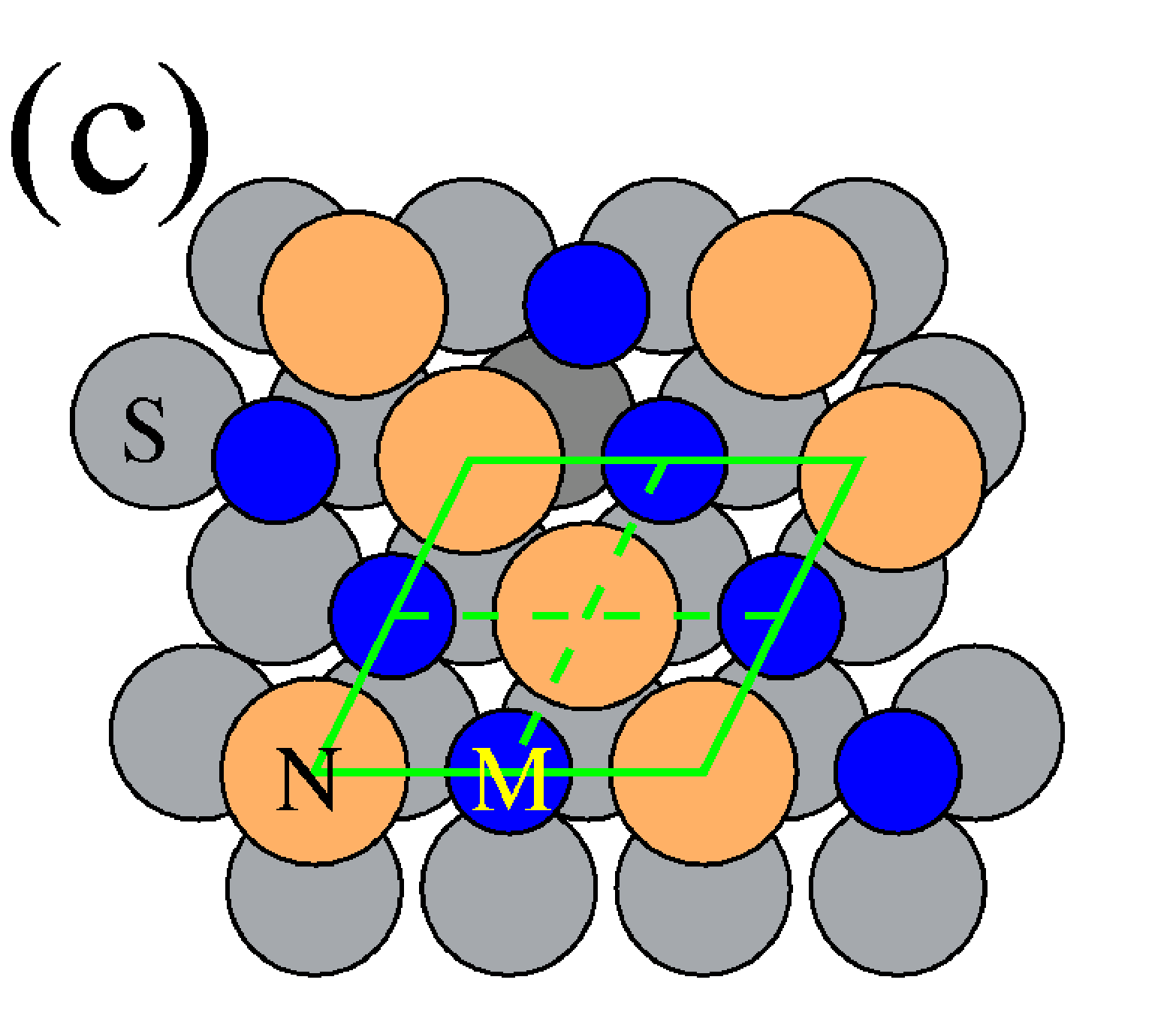}
\includegraphics[scale=0.15]{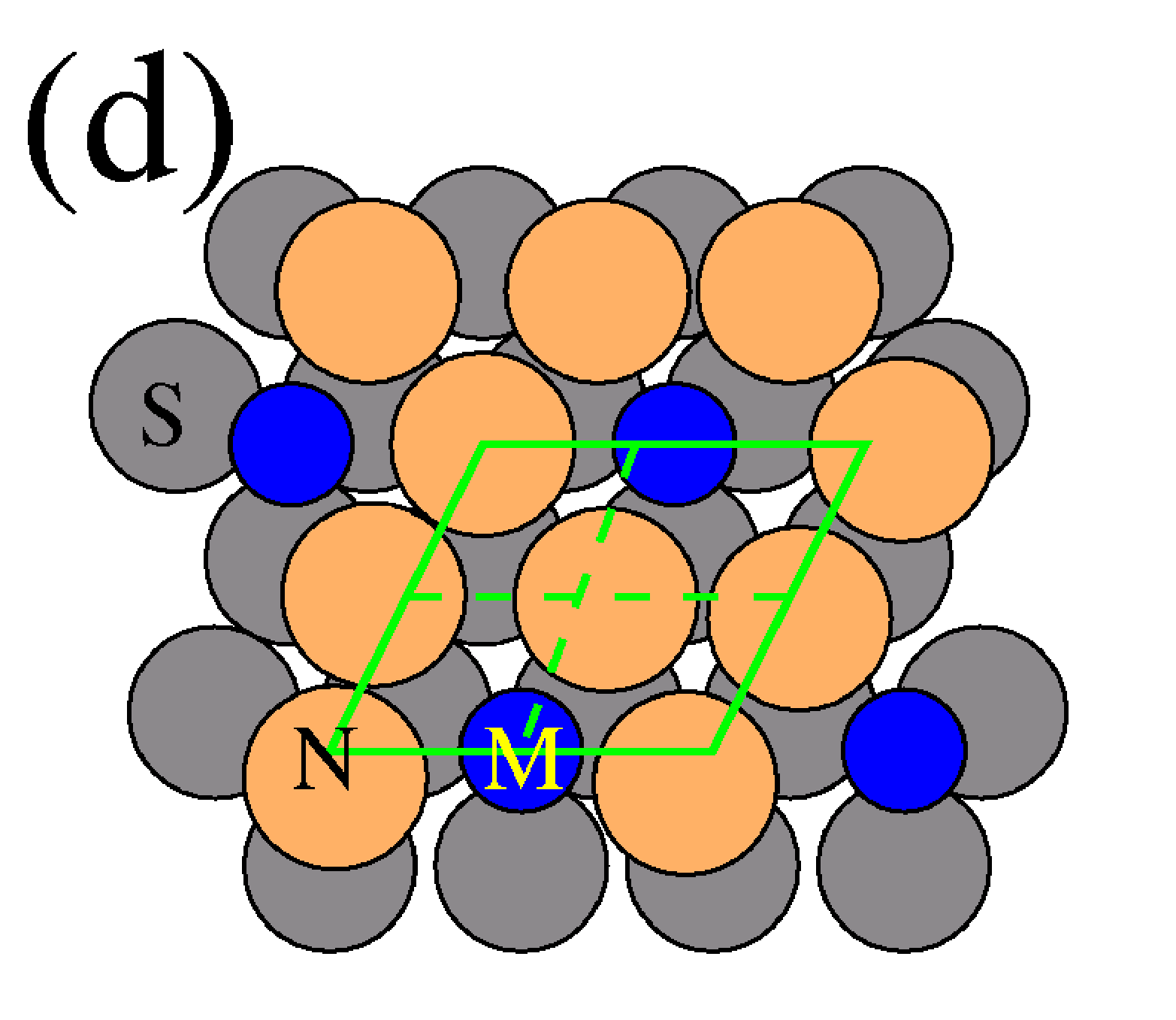}
\includegraphics[scale=0.15]{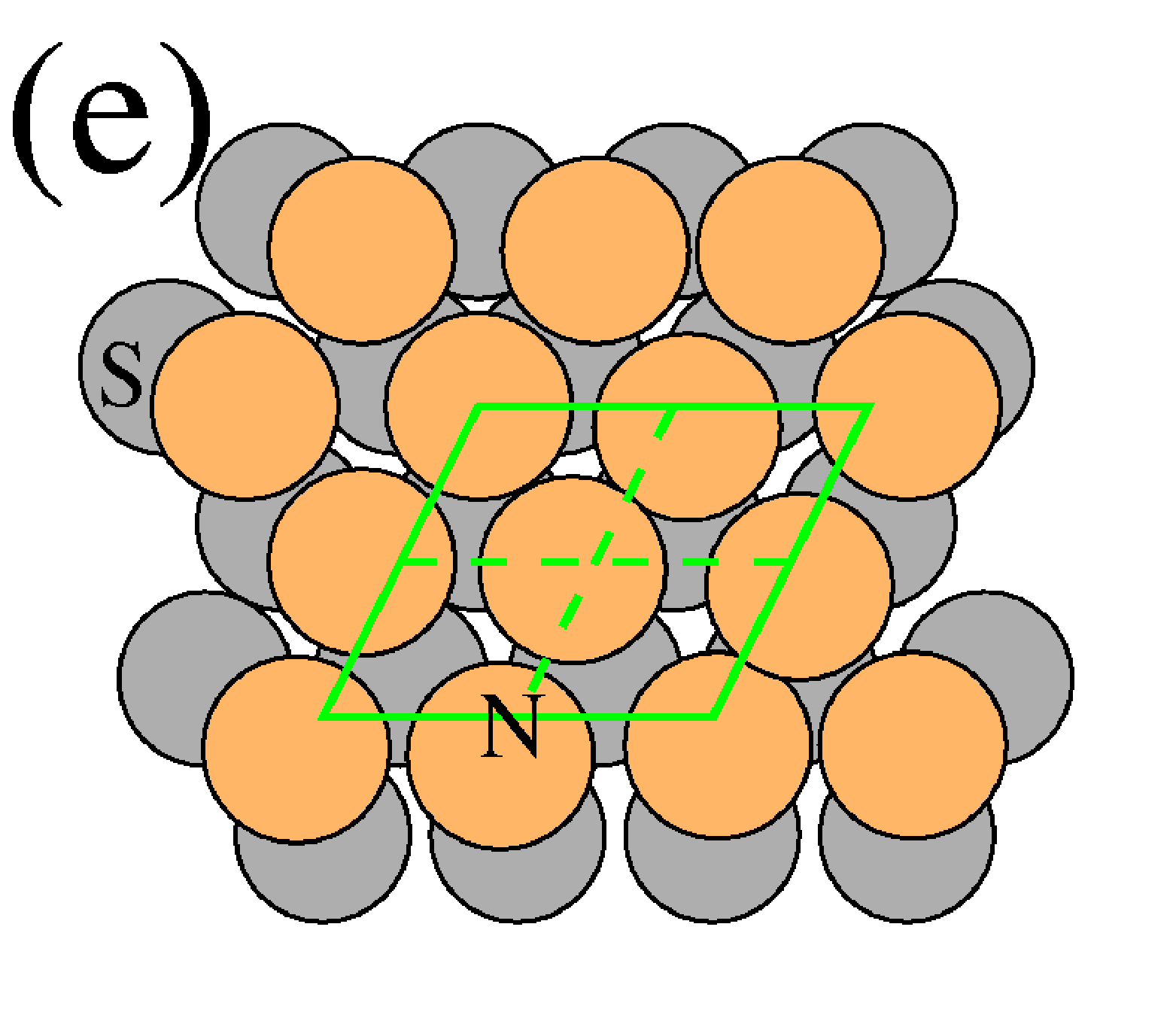}
\caption{Top views of the systems studied, for $x$ = (a) 1.00, (b) 0.75, (c) 0.50, (d) 0.25, and (e) 0.00.  
The green lines indicate the boundaries of the 2$\times$2 surface unit cell.  $S$ denotes the substrate atoms (gray) 
and  $M$ and $N$ denote magnetic (blue) and non-magnetic (orange) elements respectively.} \label{fig-topview}
\end{figure}

The (0001) surface of Ru is a closed-packed surface, on which typically one of the hollow sites, 
either hexagonal-closed packed (hcp) or face-centered cubic (fcc), is energetically preferred. 
We have allowed for both possibilities. The use of a (2$\times$2) unit cell enables us to study five different compositions as shown in Fig.~\ref{fig-topview}. 
Because of the small size of the unit cell, there is only one distinct configuration corresponding to each composition.

In this particular study, we have considered only a single, 
pseudomorphic and ordered layer of an alloy on the substrate slab. 
The general observation that reconstruction in the overlayer occurs only after a
certain critical thickness of deposited material, validates our assumption that the monolayer
(of either a single metal or an alloy) remains pseudomorphic.
For most systems studied previously consisting of a single metal
on Ru(0001), it is found that the first overlayer of the metal does not reconstruct.
However, Ag on Ru(0001) is an exception, 
in which a misfit dislocation structure has been observed, even for submonolayer films.\cite{AgonRu}

\section{Results and discussion}
The substrate element, Ru, has the hcp structure in the bulk. Upon optimizing the geometry for bulk Ru,
using the experimental
$c/a$ ratio of 1.584, we obtain $a$ (which is also the NN distance $a_S$) as 2.74 \AA, which
is close to the experimental value of 2.70 \AA.\cite{AscMer} For bulk Fe, Co, Pt, Au, Ag, Cd and Pb
we obtain NN distances $a \equiv a^{calc}_{bulk}$ of 2.47, 2.49, 2.83, 2.93, 2.95, 3.04 and 3.56 respectively.
Again, all these numbers match very well with the corresponding experimental values. 

For a single-component monolayer  on Ru(0001), we find that both the magnetic elements prefer to occupy the hcp sites; occupying instead the fcc sites costs about 75 meV per surface atom. However, for all the non-magnetic
elements, with the exception of Pt, we find that the fcc site is very slightly favored over the hcp one,
with an energy difference of the order of 4 meV per surface atom. In the rest of this paper, we work
with the structures corresponding to the favored site occupancies for each system. 

\begin{table}[tb]
\begin{center}
\begin{tabular}{|c|c|c|c|c|c|c|c|}
\hline
Elt. & Fe & Co & Pt & Au & Ag & Cd & Pb \\
\hline
$d_{12}$ (\AA) & 2.08 & 2.01 & 2.79 & 2.95 & 2.92 & 2.97 & 3.00 \\
\hline
\end{tabular}
\end{center}
\caption{Results for the value of $d_{12}$, the interplanar distance 
between the overlayer and topmost Ru layer, for single-component monolayers of $M$ or $N$ on the
Ru(0001) substrate. It is interesting to compare these results with $d_{bulk}^{Ru}$ = 2.17 \AA, which is the  
value of the interlayer distance in bulk Ru.}\label{tab-d12}
\end{table}

\begin{figure}[tb]
\centering
\includegraphics[scale=0.23,angle=270]{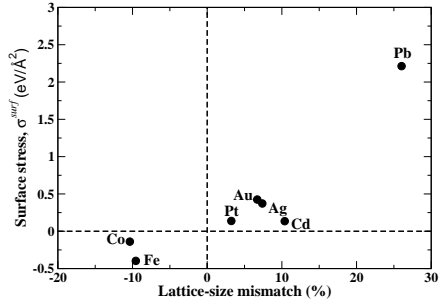}
\caption{Results for $\sigma^{surf}$, the diagonal component of the surface stress tensor, for
a single-component monolayer of either $M$ or $N$ on Ru(0001), as a function of the lattice
mismatch, defined as  $(a- a_S)/({{a + a_S}\over {2}})$.  Note for the magnetic elements, $a<a_S$, and
the surface stress is found to be tensile (negative), while for the non-magnetic elements, $a>a_S$, and
there is compressive (positive) surface stress.}\label{fig-strs}
\end{figure}

Upon depositing the single-component monolayers of either $M$ or $N$ on Ru(0001),
and relaxing the geometry, we find that
$d_{12}$, the interplanar distance between the overlayer and the topmost Ru layer, varies significantly
depending upon the element constituting the overlayer. Our results for $d_{12}$ are given in Table~\ref{tab-d12};
they may be compared with 2.17 \AA, which is the value of $d_{bulk}^{Ru}$, the interplanar distance in bulk Ru.
We see that for the magnetic elements, $d_{12} < d_{bulk}^{Ru}$, whereas for the non-magnetic elements,
$d_{12} > d_{bulk}^{Ru}$. We also see a similar pattern upon examining our results for the surface stress
of these systems (see Fig.~\ref{fig-strs}): the $M$/Ru systems are under tensile stress, whereas all
the $N$/Ru systems are under compressive stress. All these findings are consistent with the idea that
the $M$ atoms at the Ru surface would like to increase their ambient electron density, whereas the opposite
is true for the $N$ atoms; this is what one would expect from simple size considerations using the values
of $a^{calc}_{bulk}$ for all the metals.

\begin{figure}[tb]
\centering
\includegraphics[scale=0.4,angle=270]{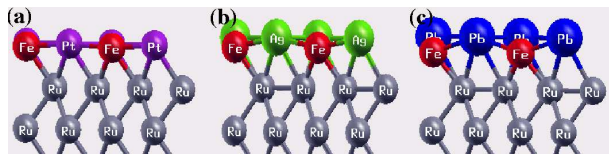}
\caption{(color online) Relaxed geometries for surface alloys with $x$ = 0.25 for (a) Fe-Pt, (b) Fe-Ag, and (c) Fe-Pb. 
Here, gray, red, purple, green and blue spheres represent Ru, Fe, Pt, Ag and Pb atoms respectively.
Note that the amount of buckling increases progressively, in keeping with what one would expect
upon considering the mismatch
between the atomic sizes of the constituent elements.}
\label{fig-buckl}
\end{figure}

For the surface alloys, we find that in every case considered by us, the hcp site is favored over the
fcc site. The difference in energy between the two sites varies from 10 to 70 meV per surface atom.
Upon relaxing the alloy structures, we find that the surface layer can exhibit significant buckling; this
follows the trends expected from the atomic-size mismatch between the constituent elements.
Thus, Pt alloys do not show any visible buckling, while  
Pb alloys show the maximum amount of buckling among all the $N$'s studied (see Fig.~\ref{fig-buckl}). 

The stability of an alloy phase relative to the phase-segregated phase can be determined by calculating 
the formation energy, $\Delta H$ which is defined as follows:
\begin{align}
\Delta H &= E_{slab}(M_xN_{1-x}/S) - xE_{slab}(M/S)\nonumber \\
         &\qquad - (1-x)E_{slab}(N/S), \label{eq-delH}
\end{align}

where $E_{slab}(A)$ is the ground state energy per surface atom for a single layer of $A$ on the substrate $S$. 
When $\Delta H$ is negative, the two metals prefer to mix rather than to segregate, 
and hence the alloy phase is more stable.

\begin{figure}[b]
\centering
\includegraphics[scale=0.25,angle=270]{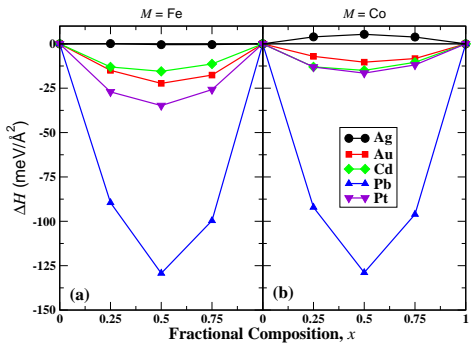}
\caption{(color online) {\it Ab initio} results for $\Delta H$, the formation energy per unit surface area: for each composition, the formation energy is plotted as 
a function of $x$, the fraction of the magnetic element. The panels on the left and right contain results for Fe and
Co alloys respectively. 
Note that 
Pb alloys are the most stable, followed by Pt, and Ag alloys are the least stable.}\label{fig-delH}
\end{figure}

Our results for $\Delta H$ as a function of a composition are presented in Fig.~\ref{fig-delH}. 
Note that in all cases, we find that $\Delta H$ is roughly symmetric about $x=0.5$, suggesting
that pairwise interactions are dominant. For both Fe and Co , alloys with Ag are found to be the
least stable and alloys with Pb appear to be the most stable. However, though Fe and Co have
almost the same $a^{calc}_{bulk}$, the values of $\Delta H$, and even the order of stability,
are not identical in the two cases.  Similarly, despite having very close values of $a^{calc}_{bulk}$,
Au and Ag display very different behavior: alloys of the former are stable, whereas 
Fe-Ag alloys are right at the boundary of stability, and
Co-Ag alloys are unstable. These observations support the view that chemical effects may, in
some cases, be quite important -- and even dominate over elastic interactions. Our finding that atomic-level
mixing is disfavored for Fe-Ag and Co-Ag is in keeping with the observations of previous authors.\cite{Thayer,
Thayer1, Yang} We point out that our results underline the fact that $(a_M + a_N)/2 \approx a_S$ need not
necessarily be a good criterion for atomic-level mixing to be favored (see Table \ref{tab-bulk-avg}).

It is generally accepted that there are two main contributions to the stability of 
such surface alloys: an elastic contribution, and a chemical contribution.\cite{Thayer1,Ozolins}
We would like to separate out the two, if possible. In order to do so, we assume that the
elastic interactions are given by a sum of NN contributions, with each pairwise term 
taking the form of a Morse potential:
\begin{equation}
V_{ij}(r) = A_0^{ij}\{1 - \exp[{-A_1^{ij}(r - b^{ij})]}\}^2\label{eq-Morse}
\end{equation}
where $r$ is the distance between the NN atoms $i$ and $j$, $b^{ij}$ is the equilibrium bond length, and
$A_0^{ij}$ and $A_1^{ij}$ are parameters related to the depth and width, respectively, of the potential well.

For each composition, the elastic energy is written as the sum of individual bond energies of the Morse form, 
by counting the total number of $M$-$M$, $N$-$N$ and $M$-$N$ bonds in each (2 $\times$ 2) unit cell.  
Accordingly, for $x$ = 0.25, 0.5 and 0.75, we obtain the elastic contribution to the formation energy as:
\begin{align}
\Delta H_{0.25}^{ela} &= 6V_{MN}(a_S) - 3V_{MM}(a_S) - 3V_{NN}(a_S)\label{eq-M0.25} \\
\Delta H_{0.50}^{ela} &= 8V_{MN}(a_S) - 4V_{MM}(a_S) - 4V_{NN}(a_S)\label{eq-M0.5} \\
\Delta H_{0.75}^{ela} &= 6V_{MN}(a_S) - 3V_{MM}(a_S) - 3V_{NN}(a_S)\label{eq-M0.75}
\end{align}

Note that Equations (\ref{eq-M0.25}) and (\ref{eq-M0.75}) are identical, i.e., within our model, the
elastic interactions lead to a $\Delta H^{ela}$ that is symmetric about $x=0.5$. It is also important to
note that for bulk alloys of $M$ and $N$, there are no terms analogous to the second and third terms on
the right-hand-sides of the above equations. Due to the presence of the substrate, these terms have to
be evaluated not at $b^{MM}$ or $b^{NN}$ (where they would lead to a zero contribution) but at the substrate
spacing $a_S$. As a result of this, one can expect mixing rules to be quite different for surface alloys
than for bulk alloys; we will return to this point further below.

\begin{figure}[b]
\centering
\includegraphics[scale=0.25,angle=270]{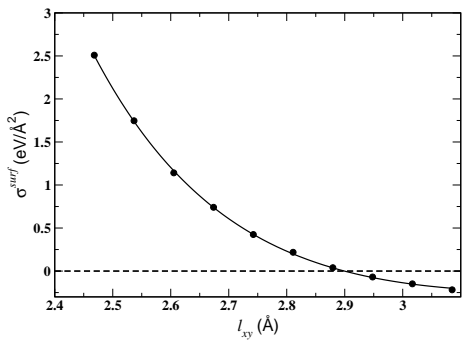}
\caption{$\sigma^{surf}$ versus $l_{xy}$ for Au/Ru(0001): The value of $\sigma^{surf}$ at each $l_{xy}$ is computed 
by compressing or stretching the complete slab, and then subtracting the contribution of substrate layers. 
The data points are fitted by an expression derived from a Morse potential, to get the values of $A_0^{Au}$, $A_1^{Au}$ and $b^{Au}$. 
The dots represent the calculated data points, while the line is the Morse fit.}\label{fig-surfstrs}
\end{figure}

In order to evaluate Eqs.~(\ref{eq-M0.25})--(\ref{eq-M0.75}), we need values for the Morse parameters $A_0$, $A_1$ and $b$, 
which appear in Eq.~\ref{eq-Morse}. We obtain these by computing the surface stress, $\sigma^{surf}$, 
for each single-component monolayer as a function of the in-plane bondlength $l_{xy}$. 
In principle, this could be obtained by compressing or expanding a monolayer of $M$ or $N$ on $S$. 
However, this would make the overlayer incommensurate with the substrate, leading to 
a surface unit cell which is too large for practical computation. 
Hence, we instead compress or expand the whole slab to perform calculations at different $l_{xy}$, 
and then subtract out the contribution from the substrate layers to the total stress, 
so as to get the surface stress at each $l_{xy}$.\cite{Pushpa} In order to carry out this procedure,
we make use of the following equation: 
\begin{align}
\sigma^{surf}(l_{xy}) &= \frac{1}{2} \bigg[(\sigma_{xx}^{V,slab})L_z - (n_a - 2)\sigma_{xx}^{V,bulk}\frac{c}{2} 
                 + \sigma_{zz}^{V,bulk}\frac{l_{xy}^2}{3c}\bigg].
\end{align}
Here, $\sigma^{surf}$ is the surface stress at an intraplanar bond length $l_{xy}$, for a slab with $n_a$ atomic layers and dimension $L_z$ (which includes the vacuum) along $z$, 
and $\sigma_{\alpha \alpha}^{V,slab}$ is the $\alpha \alpha$ component of the ``volume stress" for the
slab (i.e., it has dimensions of force per unit area, as opposed to the surface stress, which has units
of force per unit length). Similarly, $\sigma_{\alpha \alpha}^{V,bulk}$ 
is the $\alpha \alpha $-component of the volume stress for a bulk Ru cell that has been
stretched or compressed to the same $l_{xy}$ as the slab. 
Note that the geometrical factors in the last two terms are specific for the hcp structure. 

The Morse parameters for an $i$-$i$ bond can be extracted from the plot of $\sigma^{surf}$ versus $l_{xy}$ 
for each single-component overlayer of $M$ or $N$ on the Ru surface.  
As an example, our results for the variation of surface stress with in-plane strain, for a monolayer of
Au on Ru(0001), are shown in Fig.~\ref{fig-surfstrs}; qualitatively similar curves are obtained for
other elements. The value of $b^{MM}$ or $b^{NN}$ is given by the value of $l_{xy}$ at which the graph crosses
the $x$-axis, while the values of $A_0$ and $A_1$ are obtained by fitting the curve to an expression
derived from a Morse potential. The values thus obtained for all seven overlayer elements are given in Table 
\ref{tab-Morse}. The value of $b^{ii}$ serves as a measure of the effective size of an atom $i$ when placed on the Ru(0001) surface. The values obtained by us from calculations of surface stress are roughly consistent with
estimates obtained from a consideration of the buckling of the surface alloys, together with a hard-sphere model.
Note that for both the magnetic elements $M$, $b^{MM}$ is smaller than $a_S$, whereas for all the non-magnetic
elements $N$ considered by us, $b^{NN}$ is greater than $a_S$. However, the values of $b$ are found to be
different from $a^{calc}_{bulk}$, in some cases quite significantly so. This difference is due to the
presence of both the surface (i.e, no neighbors above) and the substrate (different neighbors below). It is
interesting to note that for Fe/Ru(0001), $b > a^{calc}_{bulk}$, whereas for all the other
elements, $b < a^{calc}_{bulk}$ This is presumably because Fe in the bulk form has the body centered
cubic (bcc) structure with a coordination number of 8, whereas all the other elements have either
the fcc or hcp structure with 12-fold coordination. As a result, only for Fe are the overlayer atoms
more effectively coordinated when placed on a Ru surface. However, apart from such general observations,
we were unable to discern any simple relationship connecting the values of $b$ and $a^{calc}_{bulk}$.

\begin{table}[tb]
\begin{center}
\begin{tabular}{|c|c|c|c|c|}
\hline
$M$/$N$ & $A_0$ (eV) & $A_1$ (\AA$^{-1}$) & $b$ (\AA) & $a^{calc}_{bulk}$ (\AA)\\
\hline
Fe & 0.1309 & 2.412 & 2.56 & 2.47 \\
Co & 0.5827 & 2.052 & 2.37 & 2.49 \\
\hline
Pt & 0.6744 & 1.817 & 2.79 & 2.83 \\
Au & 0.4341 & 1.797 & 2.90 & 2.93 \\
Ag & 0.3638 & 1.669 & 2.92 & 2.95 \\
Cd & 0.6564 & 1.680 & 2.79 & 3.04 \\
Pb & 0.2027 & 1.563 & 3.42 & 3.56 \\
\hline
\end{tabular}
\end{center}
\caption{Values for the Morse parameters for $M$-$M$ and $N$-$N$ interactions, as deduced from
surface stress calculations. The last column contains 
our values for the calculated nearest neighbor spacing for $M$ or $N$ in their bulk structures. These
may be compared with our values for $b$, which is the preferred intra-atomic spacing for
a monolayer of $M$ or $N$ on Ru(0001).}\label{tab-Morse}
\end{table}

It remains to obtain the Morse parameters for $M$-$N$ bonds.
In analogy with the Lorentz-Berthelot mixing rules, the $M$-$N$ bond parameters are assumed to have the form, 
$b^{MN} = (b^{MM} + b^{NN})/2$; $A_0^{MN} = \sqrt{A_0^{MM}A_0^{NN}}$ and $A_1^{MN} = \sqrt{A_1^{MM}A_1^{NN}}$.   
In Table \ref{tab-Morseavg} we have tabulated the values of $b^{MN}$. 
These should compared to the bulk Ru NN spacing (= 2.74 \AA). It is also instructive to
compare the values in Table \ref{tab-Morseavg} with those in Table \ref{tab-bulk-avg}; one finds that there
is no dramatic change upon accounting for altered surface sizes.

\begin{table}[tb]
\begin{center}
\begin{tabular}{|c|c|c|c|c|c|}
\hline
$b^{MN}$ (\AA) & Pt & Au & Ag & Cd & Pb \\
\hline
Fe & 2.67 & 2.73 & 2.74 & 2.67 & 2.99 \\
Co & 2.58 & 2.63 & 2.64 & 2.58 & 2.89 \\
\hline
\end{tabular}
\end{center}
\caption{Values of $b^{MN}$, obtained by taking the average of preferred nearest neighbor spacing on the surface; these values should be compared to our calculated value of 2.74 \AA for the NN distance on the Ru substrate.}\label{tab-Morseavg}
\end{table} 

\begin{figure}[b]
\centering
\includegraphics[scale=0.25,angle=270]{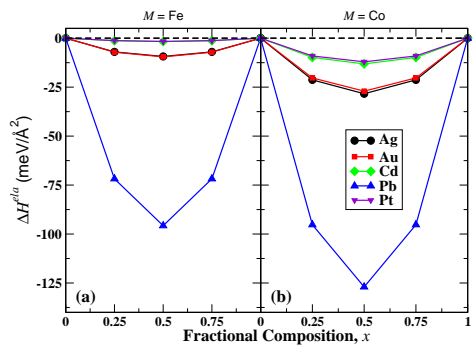}
\caption{(color online) Results for $\Delta H^{ela}$, the elastic contribution to the formation energy, as a function of fractional composition $x$, for (a) Fe and (b) Co alloys. Upon considering elastic interactions alone, Pb alloys appear to be the most stable against phase segregation, while Cd and Pt alloys are the least stable.}\label{fig-Morse}
\end{figure}

Our results for the elastic contribution to the formation energy, evaluated using Equations (\ref{eq-M0.25}) to (\ref{eq-M0.75}), are displayed in Fig.~\ref{fig-Morse}. We find that for all ten combinations considered
by us, elastic interactions always favor mixing of the two overlayer elements, in accordance with
the predictions by Tersoff.\cite{Tersoff} 

\begin{figure}[bt]
\centering
\includegraphics[scale=0.3,angle=270]{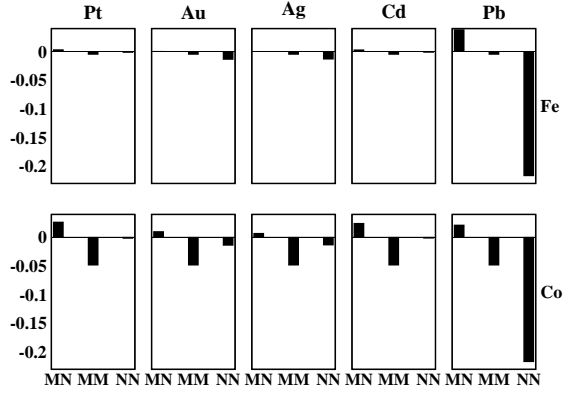}
\caption{The various contributions (from Eq.~\ref{eq-M0.5}) to the elastic contribution to the alloy
formation energy, for $x$ = 0.5. Contributions from $M$-$N$, $M$-$M$ and $N$-$N$ bonds are displayed
separately. All the histograms have been plotted 
on the same scale, to make comparison easier. Note the very high negative contribution 
from Pb-Pb bonds for both Fe-Pb (upper panel) and Co-Pb (lower panel) alloys.}\label{fig-histogram}
\end{figure}

At first sight, our most surprising result appears to be our finding that for both magnetic elements, the Pb alloys are the most stable, though upon examining Table \ref{tab-bulk-avg} or Table \ref{tab-Morseavg}, one might
think that this is unlikely. However, this is because for surface alloys, unlike bulk alloys, the phase-segregated
forms can cost a high elastic energy, because of the presence of the substrate. Since pseudomorphic Pb/Ru(0001)
costs a great deal in elastic energy, the mixed form is correspondingly favored.
In order to make this argument clearer, in Fig.~\ref{fig-histogram}, we have separated out the individual
contributions to the right-hand-side of Equation \ref{eq-M0.5}. The first ($M$-$N$) term is always positive,
while the second ($M$-$M$) and third ($N$-$N$) terms are always negative. In order for $\Delta H$ to be negative,
the first term should be small (the simple mixing rule applies only to this term), while the second and
third terms should be large in magnitude. The first term is found to follow the expectations from an elementary
consideration of sizes (either at the bulk or at the surface): Ag and Au alloys are the most favored, followed
by Pt and Cd, and then Pb. It is interesting to note that both Cd and Pt alloys have roughly the same contribution
from this first term; this is because Cd undergoes a relatively large contraction in size at the surface, relative
to the bulk. A Co monolayer on Ru(0001) is relatively unhappy (i.e., the contribution to the elastic part of
the formation energy is significant and negative), and a Pb monolayer on Ru(0001) is extremely unfavorable
energetically. As a result of these two facts, elastic interactions favor the formation of Co-$N$ alloys
over Fe-$N$ alloys, and lead to the high stability against phase segregation of $M$-Pb alloys. However, one
should be cautious in interpreting these results, since we have made the assumption that the alloys as well
as phase-segregated monolayers remain pseudomorphic. For the alloys, this is probably a valid assumption, since
the elastic energy (corresponding to the first bars in Fig.~\ref{fig-histogram}) is small, i.e., the stress is unlikely to be high enough to drive the overlayer to relax. Despite the significant elastic energy contained in a Co/Ru(0001) monolayer, it does not reconstruct.\cite{Hwang2}
However, for Pb/Ru(0001), the very high elastic energy makes it seem possible that this system might reconstruct,
presumably via a network of misfit dislocations; we are not aware of any experimental information on this system.
Thus, the high stability we obtain for $M$-Pb alloys may be misleading; the stability would be lowered if the
phase segregated form were to reconstruct (since the third term in the elastic energy would then be decreased
in magnitude).

\begin{figure}[b]
\centering
\includegraphics[scale=0.25,angle=270]{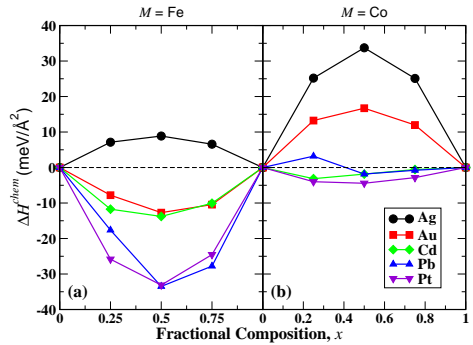}
\caption{(color online) The chemical contribution to the alloy formation energy, as a function of fractional composition
$x$, for (a) Fe and (b) Co surface alloys on Ru(0001). These graphs have been obtained by subtracting out
the elastic contribution from the {\it ab initio} results.}\label{fig-Chem}
\end{figure}

Finally, in Fig~\ref{fig-Chem}, we display our results for the chemical contribution to the formation energy,
obtained by subtracting out the data in Fig.~\ref{fig-Morse} from that in Fig.~\ref{fig-delH}.
When only chemical interactions are considered, Fe alloys are more favorable than Co alloys.
From this plot, we note that the stability of Pt alloys is largely due to the favored chemical bonds between Fe-Pt and Co-Pt ; this is consistent with the fact that these intermetallic systems also form bulk alloys.
The Ag alloys are not stable because Co-Ag and Fe-Ag bonds cost very high chemical energy, 
which cannot be offset by elastic energy; this too is consistent with previous results.\cite{Thayer, Thayer1, Yang}
Also note that Fe-Au bonds favor mixing, whereas Co-Au bonds cost energy, which explains 
the particular order of stability observed in the \textit{ab initio} results. 

\section{Summary and Conclusions}

In this paper, we have attempted to gain an understanding into the factors governing the energetics of
strain-stabilized surface alloys, by performing {\it ab initio} calculations on ten combinations involving a magnetic and a non-magnetic metal co-deposited on a Ru(0001) substrate. In many cases, we find the surface
alloy to be stable, even though the constituent elements are immiscible in the bulk.

We find that the stability (against phase-segregation) does not correlate
with expectations based upon the simple argument that the mean atomic size should be as close as possible to
the substrate lattice spacing. One reason for this is that though elastic interactions are an important mechanism governing stability, chemical interactions can also play a crucial role. In some cases, the latter are
large enough to disfavor atomic-level mixing, even if it helps in lowering the elastic energy. A second
complicating factor is that unlike for bulk alloys, for such strain-stabilized surface alloys, the phase segregated
forms can also cost elastic energy. Thus, there are three factors that determine whether or not mixing takes
place at the atomic level: (i) the elastic energy of the alloy phase, (ii) the elastic energies of the phase-segregated monolayers on the substrate, and (iii) chemical interactions. Because of this complicated situation,
a simple criterion, analogous to the first Hume-Rothery rule for bulk alloys, does not seem possible for
such systems. 

We have also found that effective atomic sizes on the Ru substrate are not equal to the bulk size; in some
cases this difference is small, while in other cases it is large. Several alloys involving a magnetic and
a non-magnetic element on a Ru(0001) surface are found to be stable against phase segregation; this is
primarily because the effective size of the magnetic elements is smaller than the nearest-neighbor distance
in the substrate, while that of the non-magnetic elements is larger, even after accounting for altered sizes
at the surface. Of the systems we have considered, we feel that Fe-Au, Fe-Cd and Co-Cd are particularly
promising candidates that would be worth experimental investigation. In these systems, both chemical and elastic interactions promote alloying. 
We have also found that surface alloys 
involving Pb and either Fe or Co appear to be very resistant to phase segregation; however, this conclusion
is dependent on our assumption that a monolayer of Pb on Ru(0001) does not reconstruct, which may or may not
be valid.   

\begin{acknowledgments}
We acknowledge helpful discussions with Sylvie Rousset, Vincent Repain and Yann Girard. Funding was provided by the Indo-French Centre for the Promotion of Advanced Research. Computational facilities were provided by the Centre for Computational Material Science at JNCASR.
\end{acknowledgments}

\end{document}